\documentclass{emulateapj}
\usepackage{color}

\begin{document}

\shortauthors{Luhman \& Hapich}
\shorttitle{Planetary-mass Brown Dwarfs in IC~348}

\title{
New Candidates for Planetary-mass Brown Dwarfs in IC~348}

\author{
K. L. Luhman\altaffilmark{1,2} and C. J. Hapich\altaffilmark{1}}

\altaffiltext{1}{Department of Astronomy and Astrophysics, The Pennsylvania
State University, University Park, PA 16802; kll207@psu.edu.}

\altaffiltext{2}{Center for Exoplanets and Habitable Worlds, The
Pennsylvania State University, University Park, PA 16802, USA}

\begin{abstract}

We have used infrared images obtained with the Wide Field Camera 3 on board the
{\it Hubble Space Telescope} to search for planetary-mass brown dwarfs
in the star-forming cluster IC~348. In those images, we have identified
12 objects that have colors indicative of spectral types later than M8,
corresponding to masses of $\lesssim30$~$M_{\rm Jup}$ at the age of IC~348.
The four brightest candidates have been observed with spectroscopy, all of
which are confirmed to have late types. Two of those candidates appear
to be young, and thus are likely members of the cluster, while the ages
and membership of the other two candidates are uncertain.
One of the former candidates is the faintest known member of IC~348 in
extinction-corrected $K_s$ and is expected to have a mass of
4--5~$M_{\rm Jup}$ based on evolutionary models and an assumed age of 3~Myr.
Four of the remaining eight candidates have ground-based photometry
that further supports their candidacy as brown dwarfs, some of which
are fainter (and potentially less massive) than the known members.

\end{abstract}

\section{Introduction}
\label{sec:intro}

Over the last decade, searches for free-floating brown dwarfs have reached
progressively deeper into the mass regime of planetary companions
($\lesssim15$~$M_{\rm Jup}$) in the solar neighborhood
\citep{cus11,luh14,kir19,bar20,mei20} and in the nearest young associations
\citep{liu13,kel15,bur16,sch16,bes17,gag18} and star-forming regions
\citep{sch12,esp17cha,zap17,lod18,esp19,rob20}.
The IC~348 cluster in the Perseus molecular cloud \citep{her08}
has been one of the most thoroughly surveyed examples of the latter
because of several characteristics:
young enough that its brown dwarfs are relatively luminous
\citep[2--6~Myr,][]{luh03,bel13};
old enough that most of its members are not heavily obscured ($A_V<3$);
among the nearest star-forming clusters ($\sim$300~pc);
sufficiently well-populated to provide good statistical constraints
on the substellar mass function ($N\sim500$);
compact enough to allow efficient imaging ($\sim0.2$~deg$^2$);
and relatively low nebular emission because of the absence of an H~II region.
The latest census of IC~348 contains 67 objects likely to be brown dwarfs
($\geq$M6.5), has a high level of completeness down to
$\sim10$~$M_{\rm Jup}$ ($\sim0.01$~$M_\odot$), and reaches masses as low as 
$\sim5$~$M_{\rm Jup}$ \citep[][references therein]{alv13,luh16,esp17}.

Brown dwarfs are typically identified using photometry or proper motions
measured from wide-field surveys or deep imaging of small fields toward
young clusters. Standard broad-band filters at optical and infrared (IR)
wavelengths have been successfully utilized in the photometric selection
process, but filters that are designed to measure absorption bands from H$_2$O
and CH$_4$ produce particularly distinctive colors for brown dwarfs
\citep{naj00,mai03,bur09,hai10,par13,tin05,tin18,all20,jos20,rob20}.
In late 2016 and early 2017, portions of IC~348 were imaged by the
{\it Hubble Space Telescope} in a medium-band filter aligned with 
H$_2$O and CH$_4$ bands and in two neighboring broad-band filters.
In this paper, we present an analysis of those data in order to search for
new members of the cluster at planetary masses.

\section{WFC3 Imaging of IC~348}

\subsection{Data Collection}
\label{sec:collect}

\begin{figure}
\epsscale{1.1}
\plotone{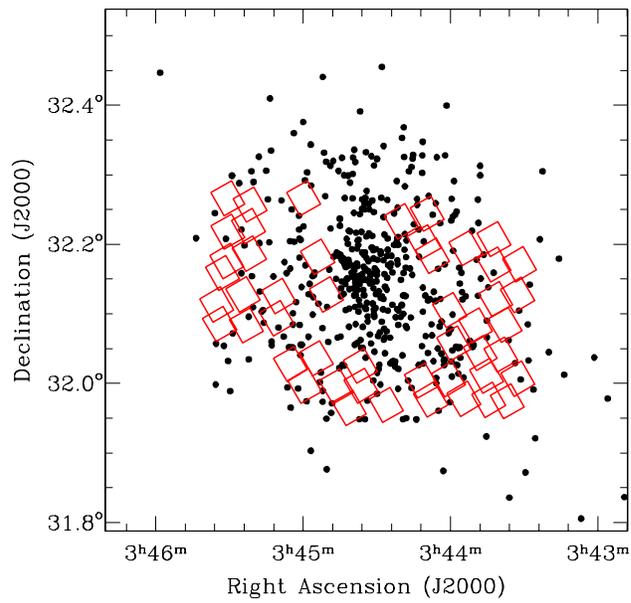}
\caption{
Map of the known members of IC~348 \citep{luh16,esp17} and the fields imaged
by WFC3 in this study.
}
\label{fig:map}
\end{figure}

IC~348 was observed with the IR channel of {\it Hubble}'s Wide Field Camera 3
\citep[WFC3,][]{kim08} on several dates between December 2016 and February 2017
through program 14626 (M. Barsony).
The camera contains a $1024\times1024$ HgCdTe array in which
the pixels have dimensions of $\sim0\farcs135\times0\farcs121$.
The inner $1014\times1014$ portion of the array detects light, which
corresponds to a field of view of $136\arcsec\times123\arcsec$.
The observations were performed with the drift-and-shift (DASH) method
of imaging multiple fields in a single orbit \citep{mom17}.
After the guide star acquisition for the initial field in a given orbit,
re-acquisitions for subsequent fields are omitted and guiding is performed
with gyros alone. During the 25~s interval between a pair of non-destructive
reads for an exposure, the telescope drift with gyros-only guiding 
is typically less than half of a WFC3 pixel, so the difference images between
adjacent reads can be shifted and combined to produce a single image that
has little smearing of the point-spread-function (PSF). For IC~348, the
total exposure times were 250 or 275~s for a given field and filter.
WFC3 observed 48 fields through three filters, consisting of
F125W (1.1--1.4~\micron), F139M (1.35--1.41~\micron), and
F160W (1.4--1.69~\micron). In a given orbit, eight fields were imaged in a 
single filter, so the data were taken across a total of 18 orbits.
The WFC3 fields are indicated on a map of the known members of IC~348 in
Figure~\ref{fig:map} \citep{luh16,esp17}.
WFC3 observed the outskirts of the cluster to avoid bright stars, which
is recommended with the DASH method \citep{mom17}.
The WFC3 fields encompass 38 of the 480 known members of the cluster.

We note that the relative exposure times among the filters were not
optimal for analysis requiring the use of all three filters (e.g., identifying
new brown dwarfs). For a given field, the same exposure time was used for
each of the three filters. The resulting sensitivity was much lower in
F139M than in the other two filters because of its smaller bandpass.
Since the F139M images measure molecular absorption from brown dwarfs,
they are essential for the photometric identification of brown dwarf candidates,
and thus the survey is effectively limited to their (more shallow) depth.

\subsection{Data Reduction}

We retrieved the raw WFC3/IR images of IC~348 from the Mikulski Archive for 
Space Telescopes: 
\dataset[https://doi.org/10.17909/t9-d358-qj35]{https://doi.org/10.17909/t9-d358-qj35}.
Each image was split into 25~s difference images using the python routine
{\tt wfc3dash}\footnote{\url{https://github.com/gbrammer/wfc3dash}}.
The resulting frames were registered and combined using the tasks
{\tt tweakreg} and {\tt astrodrizzle} within the DrizzlePac software package.
We adopted drop sizes of 1.0 native pixels and a resampled plate scale
of $0\farcs065$~pixel$^{-1}$.

For each reduced image, we used the routine {\tt starfind} in IRAF to 
identify detected sources and measure their pixel coordinates.
We aligned the world coordinate system (WCS) of each F160W image 
to astrometry of sources in the WFC3 images from Data Release 10 of the
United Kingdom Infrared Telescope Infrared Deep Sky Survey \citep{law07}. 
The WCS of each image in the other two filters was aligned to the updated WCS
for F160W.

We measured aperture photometry for the sources in the WFC3 images
using the IRAF task {\tt phot} with an aperture radius of four pixels
and radii of four and eight pixels for the inner and outer boundaries of the
sky annulus, respectively.
We measured aperture corrections between those apertures
and radii of $0\farcs4$ using bright non-saturated stars.
When using gyros-only guiding, the pointing of {\it Hubble} drifts, and
the rate of that drift varies with time.
As a result, the sampling of the PSF is not identical 
among the different fields for a given filter, and hence the aperture
corrections can vary among those fields.
Therefore, we measured an aperture correction for each image
if it contained a sufficient number of bright non-saturated stars.
For images with few stars of that kind, we applied the mean aperture
correction among all of the images for a given filter that contain bright
stars, which corresponded to 0.15, 0.16, and 0.195~mag for F125W, F139M,
and F160W, respectively. We applied those corrections and the zero-point Vega
magnitudes of 25.1439 (F125W), 23.2093 (F139M), and 24.5037 (F160W) for
$0\farcs4$ apertures\footnote{\url{https://www.stsci.edu/hst/instrumentation/wfc3/data-analysis/photometric-calibration/ir-photometric-calibration}}
to the photometry.  We adopted a minimum error of 0.02~mag for the
photometry due to uncertainties in the aperture corrections.

Some of the WFC3 fields overlap (Fig.~\ref{fig:map}), so we 
identified detections with matching coordinates among all images in a given 
filter and computed the mean coordinates and photometry for sources with 
multiple detections. The resulting catalogs for the three filters were then 
matched to each other to form a single catalog for the entire set of images.
That catalog contains 1552 sources with non-saturated detections in all three
bands.

\section{Identification of Brown Dwarf Candidates}
\label{sec:ident}

\begin{figure}
\epsscale{1.4}
\plotone{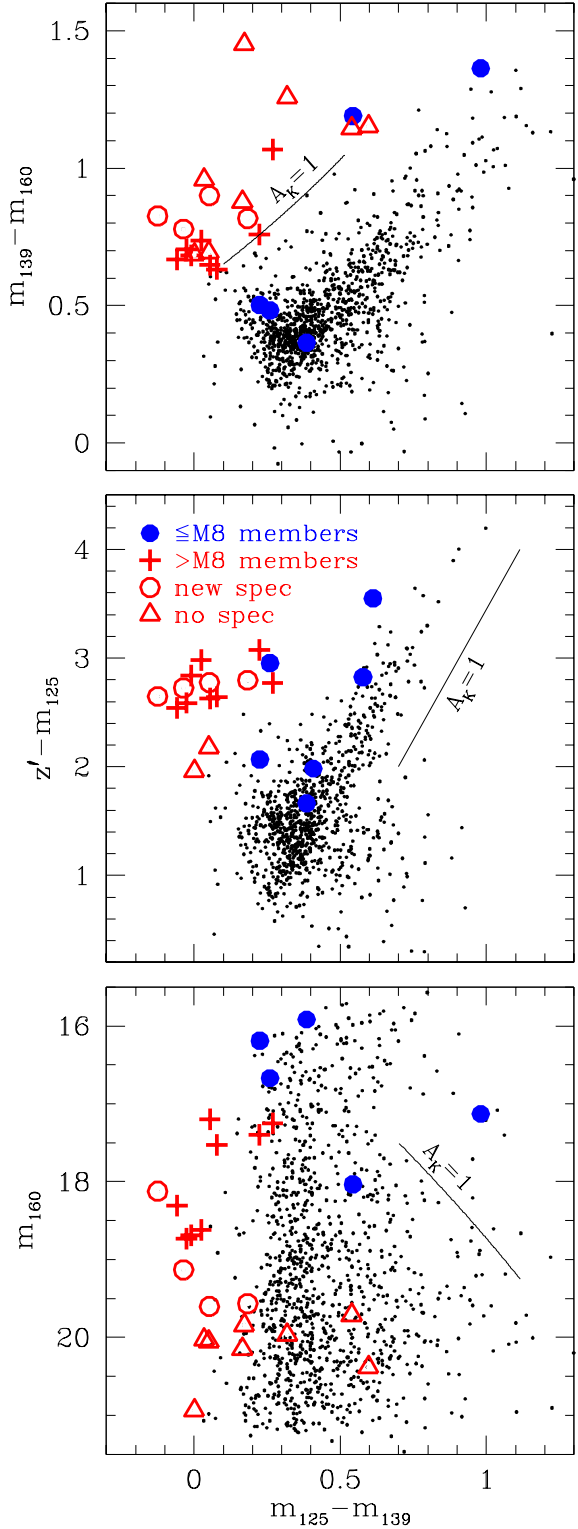}
\caption{Color-color and color-magnitude diagrams for fields in IC~348
that were imaged by WFC3, which are constructed from
three WFC3 bands and ground-based photometry in $z\arcmin$.
We have marked previously known members of IC~348 that are within the WFC3
images and are not saturated (filled circles for $\leq$M8 and crosses for
$>$M8). The $>$M8 members have distinctive positions in the two color-color
diagrams relative to most other stars. A reddening vector has been placed near
the lower envelope of the $>$M8 members in the top diagram.
We have selected candidates for new late-type members of IC~348 based on
positions above that vector (Table~\ref{tab:cand}). Four of the resulting
candidates have new spectroscopy (open circles, Fig.~\ref{fig:spec}) while 
the remaining candidates lack spectra (open triangles).
Nearly all of the sources that are not labeled as known or
candidate members of IC~348 are expected to be field stars and galaxies
(small points).
}
\label{fig:cmd1}
\end{figure}

The F139M filter of WFC3 is centered on absorption bands from H$_2$O and
CH$_4$ while F125W and F160W encompass continuum at shorter and longer
wavelengths. As a result, objects with strong absorption in those bands will
exhibit blue colors in $m_{125}-m_{139}$ and red colors in $m_{139}-m_{160}$,
which should be distinctive from most other astronomical sources.
To use those colors to identify candidate brown dwarfs in IC~348,
we have plotted a diagram of $m_{139}-m_{160}$ versus $m_{125}-m_{139}$
in Figure~\ref{fig:cmd1} for all sources from the WFC3 images that have
photometric errors less than 0.1~mag in each of three bands.
Since the WFC3 images are deeper in F125W and F160W than in F139M
(Section~\ref{sec:collect}), that sample is effectively limited by the
sensitivity in F139M and all of the selected sources have small errors
in F125W and F160W (no larger than 0.02~mag).
To further refine our selection of brown dwarf candidates, we have included
in Figure~\ref{fig:cmd1} a diagram of $z\arcmin-m_{125}$ versus
$m_{125}-m_{139}$ in which the $z\arcmin$ data were measured by \citet{luh16}
using deep imaging from \citet{alv13}.
We also show $m_{160}$ versus $m_{125}-m_{139}$ to illustrate
the range of magnitudes spanned by any candidates that we identify.
In each of the three diagrams in Figure~\ref{fig:cmd1}, a reddening vector
for $A_K=0$--1 is shown for the extinction curve from \citet{sch16av}.

In Figure~\ref{fig:cmd1}, we have indicated the known members of IC~348
that are within the WFC3 images and are not saturated. The members
earlier and later than M8 are shown with different symbols. A spectral
type of M8 is predicted to correspond to a mass of $\sim30$~$M_{\rm Jup}$
for ages of a few Myr \citep{bar98}. 
As expected, the late-type members are blue in $m_{125}-m_{139}$ and
red in $m_{139}-m_{160}$, making them distinctive from most other
sources in the WFC3 images.
To identify brown dwarf candidates based on colors of that kind, the
reddening vector in the diagram of $m_{139}-m_{160}$ versus 
$m_{125}-m_{139}$ has been placed along the lower envelope of the known
members later than M8. Twelve candidates appear above that vector, which
are marked with open circles and triangles according to whether we
have observed them with spectroscopy (Section~\ref{sec:spec}).
Six of the candidates have detections in $z\arcmin$, so they
appear in the diagram in Figure~\ref{fig:cmd1} that contains
$z\arcmin-m_{125}$. Four of those six candidates have similar positions
in that diagram as the known late-type members, which supports their
candidacy as cool objects. The other two candidates (LRL~60032 and LRL~91235)
are somewhat bluer in $z\arcmin-m_{125}$ than the known $>$M8 members.
Nine of the 12 candidates have detections in images at $J$, $H$, and $K_s$
from \citet{luh16}, so we have plotted them in a diagram of $J-H$ versus
$H-K_s$ in Figure~\ref{fig:cmd2} with all known members of IC~348.
Five of those candidates (the four with spectroscopy and LRL 61451)
have colors similar to those of the known $>$M8 members.
Two candidates, LRL 60032 and LRL 60101, depart modestly from
the known members in those colors. The two remaining candidates, LRL 52142
and LRL 60203, are significantly redder in $J-H$ than known late-type members
near the same $H-K_s$.

In the diagram of $m_{160}$ versus $m_{125}-m_{139}$ in Figure~\ref{fig:cmd1},
the 12 brown dwarf candidates range from $m_{160}\sim18$--21 and are fainter
than most of the known members of IC~348 that are within the WFC3 images.
We compare 11 of the 12 candidates to all known members of the cluster in
a diagram of $K_s$ versus $H-K_s$ in Figure~\ref{fig:cmd2}.
One of the candidates, LRL 91235, lacks photometry in those bands.
Most of the candidates are fainter than the known members in $K_s$,
although correcting the photometry for extinction would likely increase
the overlap between the candidates and known members.
If the candidates are dereddened to the intrinsic colors of
the known late-type members (e.g., $m_{125}-m_{139}\sim0$), roughly
half of the candidates would be fainter than the known members in terms
of extinction-corrected magnitudes.

We present the sample of 12 brown dwarf candidates in Table~\ref{tab:cand},
which includes coordinate-based designations, the photometry that we measured
from the WFC3 images, and photometry in $JHK_s$ from \citet{luh16}.

\section{Spectroscopy of Brown Dwarf Candidates}
\label{sec:spec}

\begin{figure}
\epsscale{0.6}
\plotone{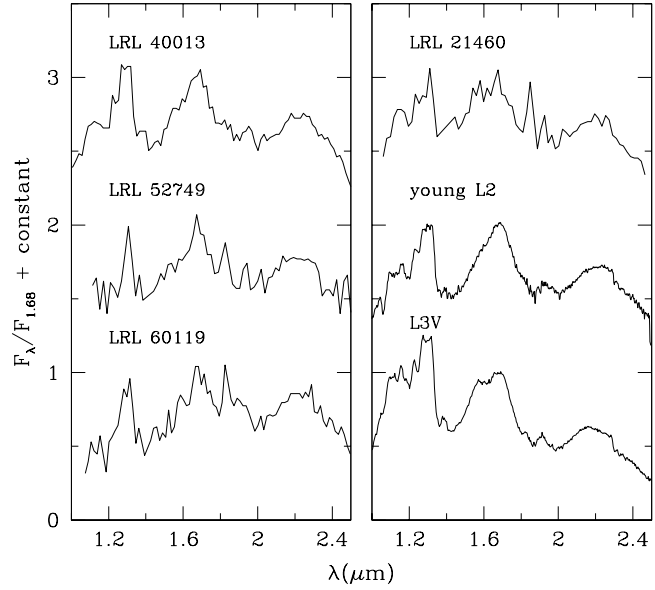}
\caption{
Near-IR spectra of four candidate substellar members of IC~348 selected
from Figure~\ref{fig:cmd1} (LRL designations), the young L2 standard spectrum
from \citet{luh17}, and a field L dwarf (2MASS J11463449+2230527).
These data are displayed at a resolution of $\sim100$.
The data used to create this figure are available.
}
\label{fig:spec}
\end{figure}

We have performed near-IR spectroscopy on the four brightest brown dwarf
candidates from the WFC3 images to measure their spectral types and check
for evidence of youth, which would support their membership in IC~348.
The spectra were obtained with the Gemini Near-Infrared Spectrograph
\citep[GNIRS,][]{eli06} during nights in February and March of 2020.
The instrument was operated in the cross-dispersed mode with the $1\arcsec$
slit and the 31.7~l~mm$^{-1}$ grating. That configuration provided a
resolution of $\sim$600 and a wavelength coverage of 0.8--2.5~\micron.
For each target, the slit was rotated to the parallactic angle and
exposures were taken at two positions along the slit separated
by $3\arcsec$ in an ABBA pattern. The numbers of exposures and exposure
times ranged from $8\times180$~sec to $12\times250$~sec.
The spectra were reduced and corrected for telluric absorption with
routines in IRAF. The reduced spectra of the four candidates are presented
in Figure~\ref{fig:spec}. The spectra have been binned to a resolution 
of $\sim100$ to improve their signal-to-noise ratios (S/N's).
The unbinned spectra are provided in an electronic file that accompanies
Figure~\ref{fig:spec}.

In Figure~\ref{fig:spec}, we have included the spectra
of a young L dwarf standard from \citet{luh17} and a field L dwarf 
standard, 2MASS J11463449+2230527 \citep[L2V,][]{kir99}. 
Like those L dwarfs, the four brown dwarf candidates exhibit 
strong absorption bands from H$_2$O, which confirms their cool nature.
The shape of the $H$-band continuum is sensitive to surface gravity and
hence age \citep{luc01}, as illustrated with the young and field L dwarf
standards in Figure~\ref{fig:spec}.
LRL~40013 and LRL~52749 have triangular $H$-band continua, indicating
that they are young, although the S/N for the latter is low enough
that we treat its age classification as tentative.
The S/N's of the spectra for LRL~60119 and LRL21460
are too low for definitive age classifications based on that feature,
although the latter may have the $H$-band plateau found in field L dwarfs.
We have measured spectral types and reddenings from the spectra of the four
candidates through comparison to the young standard spectra from \citet{luh17}.
Those classifications are listed in Table~\ref{tab:cand}.
If any candidate is a field dwarf rather than young object, then it
should be classified through a comparison to a field standard, which would
likely result in a different spectral type.

\section{Discussion}

Through spectroscopy, we have demonstrated that four of the 12 brown
dwarf candidates in IC~348 have late spectral types, two of which
appear to be young, and hence are likely to be members of the cluster.
The spectra of the other two candidates have insufficient S/N's for 
assessments of their ages and cluster membership. 
One of the candidates classified as young, LRL~52749, is the faintest
known member (by a very small margin) in extinction-corrected $K_s$.
We have estimated a mass of 4--5~$M_{\rm Jup}$ for that object
from a comparison of the luminosities predicted by evolutionary models
at an age of 3~Myr \citep{cha00,bar15} to the value derived by combining
its $K_s$ photometry with a $K$-band bolometric correction for young L dwarfs
\citep{fil15} and the distance of IC~348 \citep[321~pc,][]{ort18}.

\begin{figure}
\epsscale{1.1}
\plotone{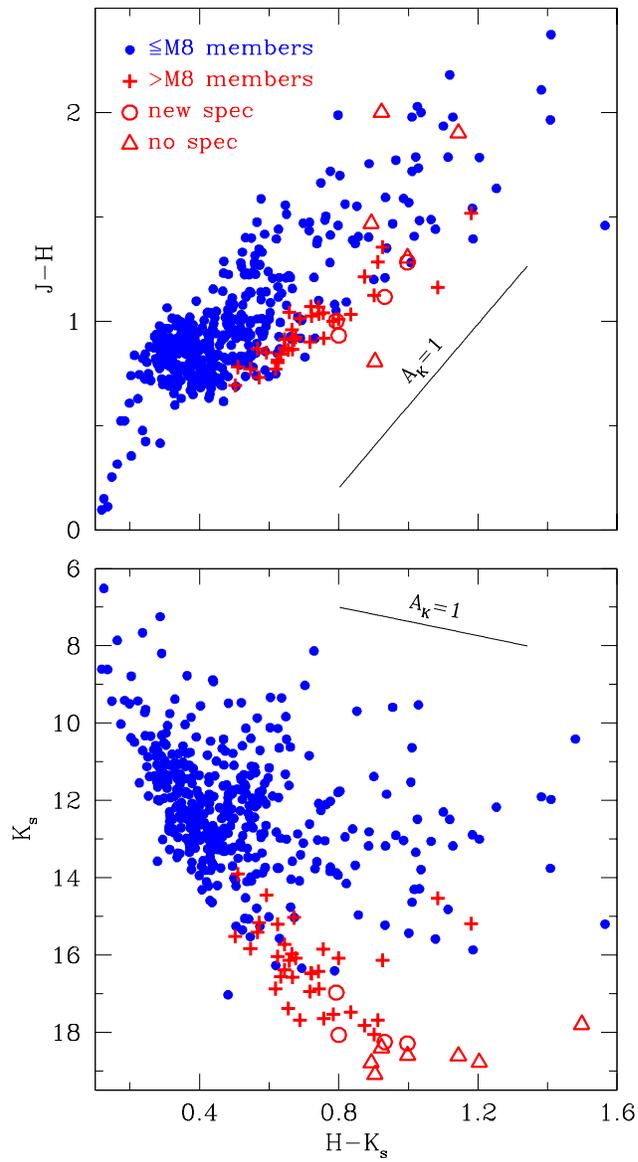}
\caption{Color-color and color-magnitude diagrams for known members of IC~348 
with spectral types of $\leq$M8 and $>$M8 (filled circles and crosses) and
candidate members selected from Figure~\ref{fig:cmd1} (Table~\ref{tab:cand})
that have new spectroscopy (open circles, Fig.~\ref{fig:spec}) and that 
lack spectra (open triangles).
}
\label{fig:cmd2}
\end{figure}

Among the remaining eight objects that lack spectroscopy,
LRL~61451 is the most promising candidate for a late-type object
based on its $J-H$ and $H-K_s$ colors (Section~\ref{sec:ident},
Fig.~\ref{fig:cmd2}). If it is a member, it could be the faintest known
member in extinction-corrected photometry.
The near-IR colors of LRL~60032 and LRL~60101 are close enough to those
of known late-type members of IC~348 that we consider them to be viable
candidates.
LRL~52542 has the reddest $H-K_s$ among the 11 candidates with measurements
of that color (Fig.~\ref{fig:cmd2}), but it is only moderately red in
$m_{125}-m_{139}$. The combination of those two colors suggests that
LRL~52542 is a highly reddened brown dwarf. If its position in the diagram
of $K_s$ versus $H-K_s$ in Figure~\ref{fig:cmd2} is dereddened to the
sequence of lightly reddened members, it would have $A_K\sim1.6$ and
$K_s\sim16.2$. LRL~60203, LRL~52142, and LRL~91235 are less likely to have
late spectral types based on $z\arcmin-m_{125}$ or $JHK_s$ colors
(Section~\ref{sec:ident}).
The last remaining candidate, LRL~61953, has too little photometry
beyond the WFC3 bands (only $H$ and $K_s$) for further assessing whether
it might have a late spectral type.

We can examine the implications of our sample of brown dwarf
candidates for the initial mass function (IMF) in IC~348.
Since the positions of the WFC3 fields were selected to avoid brighter
stars, the sample of cluster members within those fields is biased against
more massive stars. For instance, the earliest known members within
the WFC3 images have spectral types of M4 ($\sim0.3$~$M_\odot$).
As a result, the WFC3 fields cannot be used to construct an IMF that is
representative of the cluster's stellar population for the full range of
masses. Therefore, we can characterize the IMF only at lower masses in the
WFC3 fields.

\begin{figure}
\epsscale{1.1}
\plotone{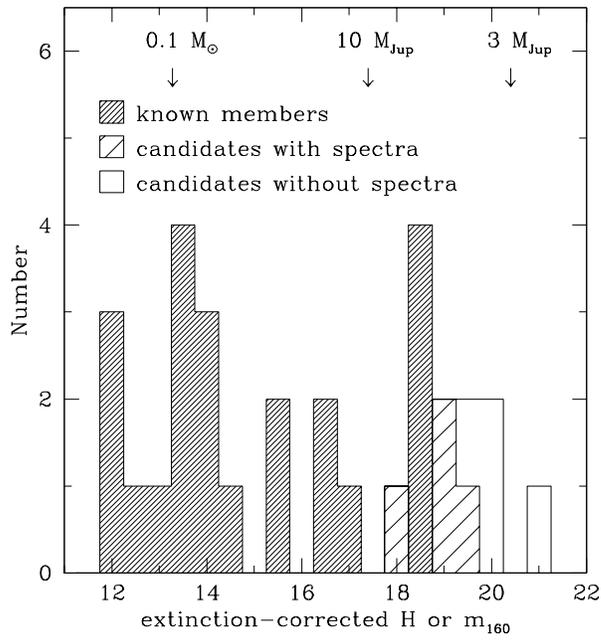}
\caption{
Distributions of extinction-corrected $H$ or $m_{160}$ for previously known
members of IC~348 that are within the WFC3 fields and have $A_K<0.6$ (finely
shaded histogram) and WFC3 candidates from Figure~\ref{fig:cmd1} that have
$A_K<0.6$. The candidates are divided into those that have been observed
spectroscopically (coarsely shaded histogram, Fig.~\ref{fig:spec}) and those
that lack spectra (open histogram). We have indicated the magnitudes that
correspond to 0.1~$M_\odot$, 10~$M_{\rm Jup}$, and 3~$M_{\rm Jup}$
for an age of 3~Myr according to evolutionary models \citep{cha00,bar15}.
}
\label{fig:histo}
\end{figure}

The current census of members of IC~348 is nearly complete for $K_s<16.8$
at $A_J<1.5$ ($A_K<0.6$) in an area that encompasses most of the WFC3
images \citep{luh16}, so we consider members and candidates within
that extinction limit for the IMF sample in the WFC3 fields.
That criterion is satisfied by 22 of the 38 known members that are within
the WFC3 images and eight of the 12 WFC3 candidates.
The four candidates that appear to have $A_K>0.6$ based on their colors
consist of LRL~52142, LRL~52542, LRL~61953, and LRL~60203.
As done in some of our previous studies of the IMF in IC~348 and other
young clusters \citep{luh16}, we use a distribution of extinction-corrected
near-IR magnitudes as an observational proxy for the IMF.
In Figure~\ref{fig:histo}, we show the distributions of extinction-corrected
magnitudes in either $H$ or $m_{160}$ for the 22 known members of IC~348 that
are within the WFC3 images and have $A_K<0.6$ and the eight
WFC3 candidates that appear to have $A_K<0.6$.
The extinctions for the known members are from \citet{luh16} and the
extinctions for the candidates are estimated from $m_{125}-m_{139}$ assuming
an intrinsic value similar to that of the known late-type members.
We use measurements of $m_{160}$ in the distributions when they are
available and otherwise use $H$ for stars that saturated in F160W.
The median value of $m_{160}-H$ is $\sim0.31$ for late-type members of IC~348,
so the two bands are sufficiently similar that they can be shown together
in the distributions for Figure~\ref{fig:histo}.

To interpret the distributions in Figure~\ref{fig:histo} in terms of
a mass function, we have marked the magnitudes that are predicted to
correspond to 0.1~$M_\odot$, 10~$M_{\rm Jup}$, and 3~$M_{\rm Jup}$
for an age of 3~Myr according to evolutionary models \citep{cha00,bar15}.
Based on those theoretical magnitudes, 
Figure~\ref{fig:histo} contains 13 objects between 0.1~$M_\odot$
and 10~$M_{\rm Jup}$, all of which are known members.
If the substellar mass function is flat in logarithmic units
($dN/dlog M\propto M^{-\Gamma}$ where $\Gamma=0$), the mass interval
from 3--10~$M_{\rm Jup}$ would contain $\sim6.5$ objects.
The distributions in Figure~\ref{fig:histo} contain four known members
and seven candidates in the magnitude range corresponding to
3--10~$M_{\rm Jup}$, and two of those candidates have been classified
as members through our spectroscopy. Thus, the mass function in the WFC3
fields is consistent with $\Gamma\gtrsim0$, although the uncertainties
in the slope are large given the small numbers objects in question
and the unknown membership status of some of the candidates.
Constraints on the mass function at planetary masses in IC~348 would
be improved through spectroscopy of the remaining candidates 
and a survey of areas of the cluster that contain larger numbers of members.

\acknowledgements

This work was supported by NASA grant 80NSSC18K0444
and is based on observations made with the NASA/ESA
{\it Hubble Space Telescope} obtained at the Space Telescope Science Institute,
which is operated by the Association of Universities for Research in
Astronomy, Inc., under NASA contract NAS 5-26555.
These observations are associated with program 14626.
The Gemini data were obtained through program GN-2020A-FT-102.
Gemini Observatory is operated by AURA under a cooperative agreement with
the NSF on behalf of the Gemini partnership: the NSF (United States), the NRC
(Canada), CONICYT (Chile), the ARC (Australia),
Minist\'{e}rio da Ci\^{e}ncia, Tecnologia e Inova\c{c}\~{a}o (Brazil) and
Ministerio de Ciencia, Tecnolog\'{i}a e Innovaci\'{o}n Productiva (Argentina).
The Center for Exoplanets and Habitable Worlds is supported by the
Pennsylvania State University, the Eberly College of Science, and the
Pennsylvania Space Grant Consortium.

\clearpage

\begin{deluxetable}{llllllllll}
\tabletypesize{\scriptsize}
\tablewidth{0pt}
\tablecaption{Candidate Members of IC~348\label{tab:cand}}
\tablehead{
\colhead{IC 348 IRS} &
\colhead{LRL\tablenotemark{a}} &
\colhead{Spectral} &
\colhead{Young?} &
\colhead{$m_{125}$} &
\colhead{$m_{139}$} &
\colhead{$m_{160}$} &
\colhead{$J$} &
\colhead{$H$} &
\colhead{$K_s$}\\
\colhead{} &
\colhead{} &
\colhead{Type/$A_K$} &
\colhead{} &
\colhead{(mag)} &
\colhead{(mag)} &
\colhead{(mag)} &
\colhead{(mag)} &
\colhead{(mag)} &
\colhead{(mag)} 
}
\startdata
J03433364+3201037 & 61953 & \nodata & ? & 22.13$\pm$0.03 & 21.53$\pm$0.09 & 20.38$\pm$0.02 & \nodata & 19.98$\pm$0.10 & 18.78$\pm$0.06 \\
J03433755+3201489 & 21460 & M9--L2/0.15 & N? & 19.87$\pm$0.02 & 19.91$\pm$0.02 & 19.13$\pm$0.02 & 19.80$\pm$0.02 & 18.87$\pm$0.04 & 18.07$\pm$0.03 \\
J03433807+3208133 & 60101 & \nodata & ? & 21.19$\pm$0.02 & 21.02$\pm$0.05 & 20.15$\pm$0.02 & 21.15$\pm$0.09 & 19.68$\pm$0.07 & 18.79$\pm$0.04 \\
J03433941+3208131 & 60032 & \nodata & ? & 20.80$\pm$0.02 & 20.75$\pm$0.04 & 20.05$\pm$0.02 & 20.80$\pm$0.05 & 20.00$\pm$0.12 & 19.09$\pm$0.05 \\
J03434218+3212130 & 91235 & \nodata & ? & 21.62$\pm$0.02 & 21.62$\pm$0.08 & 20.94$\pm$0.02 & \nodata & \nodata & \nodata \\
J03434453+3209113 & 60119 & L0--L4/0.20 & ? & 20.56$\pm$0.02 & 20.50$\pm$0.03 & 19.60$\pm$0.02 & 20.57$\pm$0.03 & 19.29$\pm$0.03 & 18.29$\pm$0.02 \\
J03435236+3158556 & 61451 & \nodata & ? & 21.02$\pm$0.02 & 20.99$\pm$0.04 & 20.03$\pm$0.02 & 20.91$\pm$0.09 & 19.60$\pm$0.06 & 18.60$\pm$0.02 \\
J03441321+3200588 & 52542 & \nodata & ? & 21.46$\pm$0.02 & 21.29$\pm$0.07 & 19.84$\pm$0.02 & \nodata & 19.30$\pm$0.05 & 17.80$\pm$0.02 \\
J03442211+3214105 & 40013 & M9.5--L2/0.10 & Y & 18.82$\pm$0.02 & 18.95$\pm$0.02 & 18.12$\pm$0.02 & 18.77$\pm$0.02 & 17.77$\pm$0.02 & 16.98$\pm$0.02 \\
J03443946+3156549 & 60203 & \nodata & ? & 21.40$\pm$0.02 & 20.86$\pm$0.03 & 19.71$\pm$0.02 & 21.33$\pm$0.08 & 19.33$\pm$0.03 & 18.40$\pm$0.03 \\
J03444366+3158452 & 52142 & \nodata & ? & 21.55$\pm$0.02 & 21.23$\pm$0.05 & 19.97$\pm$0.02 & 21.66$\pm$0.12 & 19.76$\pm$0.05 & 18.62$\pm$0.02 \\
J03445439+3209485 & 52749 & L0--L4/0.10 & Y? & 20.57$\pm$0.02 & 20.38$\pm$0.04 & 19.57$\pm$0.02 & 20.30$\pm$0.02 & 19.18$\pm$0.03 & 18.25$\pm$0.02 
\enddata
\tablenotetext{a}{These source names are a continuation of the
designation numbers from \citet{luh98}.}
\end{deluxetable}

\clearpage

\end{document}